\documentclass[useASM]{mn2e}
\usepackage{graphicx}
\usepackage{ulem}

\title[]
{Constraining Einstein's Equivalence Principle With Multi-Wavelength Polarized astrophysical Sources}

\author[Yi et al.]
{Shuang-Xi Yi$^{1}$\thanks{yisx2015@qfnu.edu.cn}, Yuan-Chuan Zou$^{2}$, Jun-Jie Wei$^{3,4}$ and Qi-Qi Zhou$^{1}$\\
 $^1$School of Physics and Physical Engineering, Qufu Normal University, Qufu 273165, China\\
 $^2$School of Physics, Huazhong University of Science and Technology, Wuhan 430074, China\\
 $^3$Purple Mountain Observatory, Chinese Academy of Sciences, Nanjing, 210023, China \\
 $^4$School of Astronomy and Space Sciences, University of Science and Technology of China, Hefei 230026, China\\}

\begin{document}

\maketitle

\label{firstpage}
\begin{abstract}
The observed time delays between photons with different circular polarizations from an astrophysical object provide a new, interesting way of
testing the Einstein Equivalence Principle (EEP). In this paper, we constrain the EEP by considering both Shapiro time delay and Faraday rotation effects. We continue to search for astronomical sources that are suitable for testing the EEP accuracy, and obtain 60 extragalactic radio sources with multi-wavelength polarization angles in three different radio bands (20, 8.6, and 4.8 GHz)
and 29 brightest stars within our own Milky Way galaxy with multi-colour linear polarimetric data in five optical bands ($UBVRI$). We apply the Metropolis-Hastings Markov Chain to
simulate the fit parameters. The final results show that the values of the parameterized post-Newtonian parameter $\gamma$ discrepancy ($\Delta \gamma_{p}$) are constrained to be in the range of $10^{-26}-10^{-23}$ for 60 radio sources and in the range of $10^{-23}-10^{-20}$ for 29 optical polarization stars. Compared to previous EEP tests that based on the single polarization measurement in the gamma-ray band, our results have profound superiority that nearly a few tens of astrophysical sources with multi-wavelength polarization observations commonly in the optical and radio bands are available. It ensures that these sources can give more significantly robust bounds on the EEP.
Although the presented method is straightforward, the resulting constraints on the EEP should be taken as upper limits as other more complex astrophysical effects affecting a polarization rotation are hardly considered.
\end{abstract}

\section{Introduction}
The Einstein Equivalence Principle (EEP) is one of the cornerstones of many metric theories of gravity, including general relativity. According to the EEP, the test particle traveling in the same gravitational potential
in vacuum is independent of its internal structure, energy, or composition. Many metric theories of gravity satisfying the EEP predict that the parameterized post-Newtonian (PPN) parameter $\gamma$ of two different test particles should be the same, i.e., $\gamma_a=\gamma_b\equiv\gamma$, where the subscripts denote two different particles (Will 2006, 2014; Gao et al. 2015; Minazzoli et al. 2019). Therefore, two test particles emitting from the same source to the observer should have the same time delay. The accuracy of the EEP can then be constrained with the observed time delays for different types of messenger particles (e.g., photons, neutrinos, or gravitational waves), or the same type of particles but with different energies or different polarization states (Shapiro 1964; Wu et al. 2017; Yang et al. 2017; Wei \& Wu 2019).

In the literature, the arrival time differences of different messenger particles have been applied to test the EEP through the relative differential variations of the $\gamma$ values, such as the particle emissions from SN1987A (Krauss \& Tremaine 1988; Longo 1988), gamma-ray bursts (GRBs) (Gao et al. 2015; Yu et al. 2018; Yang et al. 2017), fast radio bursts (FRBs) (Wei et al. 2015; Nusser 2016), blazars (Wang et al. 2016; Wei et al. 2016), the Crab pulsar (Yang \& Zhang 2016; Zhang \& Gong 2017), and gravitational wave sources (Wu et al. 2016; Kahya 2016). The resulting constraints on the EEP accuracy have been improved by several orders of magnitude using these astrophysical objects.

Since the polarization is supposed as a basic component of the internal structure of photons, the polarization measurements from astrophysical objects can also serve as an ideal test bed to probe the EEP (Yang et al. 2017; Wu et al. 2017; Wei \& Wu 2019; Yi et al. 2020). Yang et al. (2017) took GRB 110721A with high linear polarization in the gamma-ray band as an example to constrain the accuracy of the EEP. They obtained the current best constraint on the EEP of $\Delta \gamma_{p} < 1.6\times10^{-27}$. Recently, Yi et al. (2020) provided a new method to test the EEP with multi-wavelength radio observations of polarized blazars. They selected 15 groups of polarization data from the blazar 3C 279, and obtained an upper limit of $\Delta \gamma_{p} = (1.91\pm0.34)\times10^{-20}$ when considering both the Shapiro time delay and the simple form of Faraday rotation effects.
However, the number of available polarization measurements in the gamma-ray band for this particular kind of test
is very limited. Therefore, the outcomes of EEP lack significant statistical robustness even though some upper limits of $\Delta \gamma_{p} $ were set to be extremely small (Yang et al. 2017; Wei \& Wu 2019). And fortunately, many astrophysical sources with multi-wavelength polarization measurements at radio and optical bands have been observed. It is always helpful to constrain the EEP using different methods and different data.
In this work, we gather the multi-wavelength polarization observations of 60 radio sources and 29 brightest stars to test the EEP through a novel method which was discussed in Yi et al. (2020). The number of these astrophysical sources with polarization measurements overwhelms the size of data sets of previous EEP tests and guarantees much more significant statistical robustness.

\section{Method Description}

Considering the Shapiro time delay effect, the time interval required for test particles to
traverse a given distance would be longer by (Shapiro 1964; Krauss \& Tremaine 1988; Longo 1988):
\begin{equation}
\delta t_{\rm gra} = -\frac{1+\gamma}{c^{3}}\int_{\bf r_{\rm e}}^{\bf r_{\rm o}}U( r) d r,
\end{equation}
where $ r_{\rm o}$ and $ r_{\rm e}$ represent locations of Earth and source, respectively, $U( r)$ is the
gravitational potential. We consider a linearly polarized light, which is a superposition of two monochromatic waves with opposite circular polarizations (labeled with `r' and `l'). Once the EEP fails,
photons with right- and left-handed circular polarizations radiated simultaneously
from the source will arrive at the Earth with a time delay difference. It should be underlined that this is true
only in the case of gravitational theories in which the antisymmetric part of the metric tensor is coupled with the electromagnetic field. The relative Shapiro time delay is given by
\begin{equation}\label{2}
\Delta t_{\rm gra} = \mid\frac{\Delta \gamma_{\rm p}}{c^{3}}\int_{\bf r_{\rm e}}^{\bf r_{\rm o}}U(r) d r\mid,
\end{equation}
where $\Delta \gamma_{\rm p} \equiv \gamma_{\rm l}-\gamma_{\rm r}$ is the difference of the $\gamma$ values for
different circular polarization states. To estimate the accuracy of the EEP in Eq. (2), one has to figure out the
gravitational potential along the propagation path and the Shapiro time delay.

For a cosmological source, Nusser (2016) first pointed out that the potential fluctuations from the large-scale structure are much larger than the gravitational potential of the Milky Way. They proved that incorporating
the gravitational potential from the large-scale structure could tighten the EEP constraints by about four
orders of magnitude. Laniakea is a newly discovered supercluster of galaxies, which is the closest and most massive gravitational
body to our Milky Way galaxy (Tully et al. 2014). In the previous works, the Laniakea supercluster of galaxies
has been used as the deflector in the Shapiro delay tests (Luo et al. 2016; Wei \& Wu 2019).
Here we also adopt the Laniakea's gravitational potential to calculate the Shapiro delay.
Considering a cosmological source, Laniakea could be regarded as a point-mass approximation when estimating the gravitational potential.
We use a Keplerian potential $U(r)=-GM/R$ here for Laniakea, thus we have (Longo 1988)
\begin{equation}\label{eq:gammadiff}
\Delta t_{\rm gra}= \Delta\gamma_{p} \frac{GM_{\rm L}}{c^{3}}
\ln \left\{ \frac{ \left[d+\left(d^{2}-b^{2}\right)^{1/2}\right] \left[r_{L}+s_{\rm n}\left(r_{L}^{2}-b^{2}\right)^{1/2}\right] }{b^{2}} \right\}\;,
\end{equation}
where $M_{\rm L}\simeq 1\times10^{17}M_{\odot}$ is the Laniakea mass (Tully et al. 2014), $G=6.68\times10^{-8}$ erg g cm$^{-2}$ is the gravitational value, $c=3\times10^{10}$ cm s$^{-1}$ is the speed of light, $d$ is the distance
from the source to our Earth, $b$ represents the impact parameter of the light paths relative to the Laniakea center,
$r_{L}\simeq77$ Mpc denotes the distance from the Laniakea center to the Earth (Lynden-Bell et al. 1988),
and $s_{\rm n}=+1$ or $s_{\rm n}=-1$ correspond to the cases where the source is located along the Laniakea or anti-Laniakea center.
More details about the gravitational potential from the large-scale structure for a cosmological source can be seen in Nusser (2016), Luo et al. (2016) and Wei \& Wu (2019),  while the gravitational potential U(r) for an object in the Milky Way galaxy can be seen in Zhang \& Gong (2017).

Considering the relative Shapiro time delay $\Delta t_{\rm gra}$ discussed above,
the rotation of the linear polarization angle $\Delta \phi$ during the traverse from the source to the observer is given by (Yi et al. 2020):
\begin{equation}
\Delta \phi =   \frac{ \pi c \Delta t_{\rm gra}}{\lambda},
\end{equation}
where $\lambda$ is the wavelength. However, there are also existing some astrophysical effects
that modify the linear polarization angle of radio sources, such as Faraday rotation, when the light goes through the magnetized plasmas.
The linear polarization angle is consistent with the $\lambda^2$ law expected for Faraday rotation with the simple formalism\footnote{The form of $\lambda^2$ for the Faraday Rotation effect used in our paper is only the simple model, the real condition may have more complex wavelength dependence of the polarization angle rotation, but the multiple rotation measure components are hardly confirmed.} (e.g. Burn 1966; Hovatta et al. 2019).
Therefore, the observed linear polarization angle at wavelength $\lambda$ emitted from an astronomical event could
be regarded as the new expression when considering both the EEP and Faraday rotation effects simultaneously,
\begin{equation}\label{eq:6}
\phi_{\rm obs}( \lambda) = \phi_{\rm 0}+\frac{A}{\lambda}+B\lambda^2,
\end{equation}
where $\phi_{\rm 0}$ is the initial angle of the linearly polarized light, $A \equiv { \pi c} \Delta t_{\rm gra} $ denotes the contribution from the Shapiro time delay effect, and $B$ is the rotation measure induced by Faraday rotation (Burn 1966; Hovatta et al. 2019).
$\phi_0$, $A$, and $B$ are free parameters that can be optimized from the fit to the observational
data, if there are linear polarization measurements in several bands from the astronomical object.
For more details please see Yi et al. (2020).

We describe the method of testing EEP with multi-wavelength observations of polarized astronomical sources above.
This method is simple but with high accuracy of the EEP test. We acquire the value of $\Delta {\gamma}_{\rm p}$ with the time lag $\Delta t_{\rm gra}$ ($\Delta t_{\rm gra}=A/\pi c$),
which is obtained by the fit of the observational data.
Although the presented method is straightforward, some potential effects, especially the opacity effects in
the radio window have not been taken into account, such as differential Faraday rotation (O'Sullivan et al. 2012; Karamanavis et al. 2016; Pasetto et al. 2016), and hence the constraint results on the EEP should be still taken as upper limits. Therefore, considering the opacity effects for radio bands, we also search some optical polarization sources, which should be much less affected by the Faraday rotation ($\propto \lambda^2$).

\section{Tests of the EEP using Polarized Radio and Optical Sources}
It is feasible to select one or two suitable candidates for testing the EEP with the new method. However, it makes more sense to take more polarized samples for constraining the upper limits of $\Delta{\gamma}$.
We can get a limited range of $\Delta{\gamma}$ when using the catalogue of different polarized objects at different distances.
Therefore, it is necessary to constantly test the validity of the EEP with more appropriate polarized astronomical sources through the new constraint method. We continue to search for such astronomical sources and collect the observations of the Australia Telescope 20-GHz (AT20G) Survey with multi-wavelength polarization data. According to the introduction of Massardi et al. (2008), the full survey of the AT20G covers the whole southern sky to a flux density limit of $\simeq$ 50 mJy, and the project began in 2004 and ended in 2007. More than 4400 sources were obtained, but they only presented the analysis of the brightest sources with the flux $S_{20~ GHz} > 0.50 ~Jy$ in the AT20G Survey (Massardi et al. 2008). Follow-up polarization measurements at 20, 8.6, and 4.8 GHz were carried out for those radio sources. However, according to Massardi et al. (2008), some polarization data were not observed at some bands (20, 8.6, and 4.8 GHz) (see their Tables 2 and 3). Only about 60 radio sources are collected with the multi-wavelength polarized data. Table 1
lists 60 selected radio sources with different positions, redshifts, and polarization angles (with no errors, please also see Massardi et al. 2008) in three bands, respectively.

According to Feinstein et al. (2008), we also present the multi-colour linear polarimetric data for 29 of the brightest stars in the area of the open cluster NGC 6250 in the Milky Way galaxy. The age of NGC 6250 could be estimated as $1.4\times10^7$ yr with a distance of about 1025 pc to the earth ($l=340^{\circ}.8, b=-1^{\circ}.8$). Data on linear optical polarimetry were obtained during 2004 and 2005 in Argentina. Each optical polarized star was observed simultaneously through the Johnson-Cousins broad-band $UBVRI$ filters. The multi-colour polarimetric observations are listed in Table 2.

We compile those polarization observations from the astrophysical sources, and obtain the wavelengths and the corresponding polarization angles.
We suppose that the wavelength $\lambda$ is an independent variable, and the observed polarization angle $\phi_{\rm obs}$ is a dependent variable. For the nonlinear regression about the polarized observations of each source, we use the function $\phi$ ($\lambda$) to reproduce the polarized data. We then apply the Metropolis-Hastings Markov Chain Monte Carlo (MHMC) method to simulate the linear polarization angles and the corresponding wavelengths. The observed data of each group is simulated with three unknown parameters, e.g., $A$, $B$, and $\phi_0$. Figure 1 shows two cases of 60 radio sources with the observed linear polarization angles at different wavelengths, and the red line is the best fit line with Eq. (5). Figure 2 presents two cases of 29 optical stars with the polarization observations. Using this fitting method, we obtain the best-fit results for those sources with the multi-wavelength linear polarized data, which are listed in Tables 3 and 4.

The Shapiro time delay between photons with different circular polarizations for each source can be estimated with the optimized value of $A$, i.e., $\Delta t_{\rm gra} = \frac{A}{ \pi c}$. We find that the estimated Shapiro time delays are in the range of $10^{-13}-10^{-11}$ s for the radio sources and in the range of $10^{-18}-10^{-15}$ s for the optical candidates. Taking the relevant observation information into Eq. (3), such as the position and distance of the source, we then obtain the upper limits for $\Delta \gamma_{p}$. The constraints on the $\gamma$ discrepancy for each source are shown in Tables 3 and 4. We find the constraint results of $\Delta \gamma_{p}$ are in the range of $10^{-26}-10^{-23}$ for 60 radio sources and in the range of $10^{-23}-10^{-20}$ for the optical polarized stars, which are one of the best limits on the EEP through the new method with the multi-wavelength polarization observations of astrophysical sources. Our constraint results with the radio polarized sources are an improvement of 5-6 orders of magnitude over some previous analysis when considering the large scale structure potential. The distributions of the constraint results on EEP with radio and optical polarized candidates are shown in Figure 3. Since some potential effects have not been taken into account,
the constraint results on the EEP should be still taken as upper limits. The specific range of $\Delta \gamma_p$ in this paper is only the
distribution of the constraint results on the EEP for each polarized source.

In this paper, we constrain the EEP by considering both Shapiro time delay and Faraday rotation
effects. We should indicate that although the presented method used in this paper can provide severe
tests of the EEP, which is also one of the cornerstones of general relativity, it cannot be applied
directly to distinguish between some specific gravity theories, for example, general relativity theory
and its alternatives. Our purpose is designed to improve the constraint results on EEP with different
polarized astrophysical sources. Thus, in order to distinguish between general relativity theory and
other alternative gravity theories, one should investigate further for developing more accurate tests
of the EEP and some related theories.

\section{Summary and discussion}
Since the polarization from astrophysical objects is supposed to be part of the internal structure of photons, we have considered that polarization measurements of astrophysical objects can provide good tests of the EEP. In other words, the validity of the EEP can be tested with the Shapiro time delay of particles with different polarization states. Unlike previous methods given by other authors, we apply our method to test EEP with multi-wavelength polarization observations of the sources by considering both the EEP and Faraday rotation effects.  Although it is straightforward to constrain the EEP based on the method, the constraint results should be taken as upper limits as other potential astrophysical effects affecting the polarization rotation are hardly considered.

We try to search for astronomical sources that are suitable for testing the EEP accuracy. Sixty radio sources with multi-wavelength polarization angles in three bands (20, 8.6, and 4.8 GHz) and twenty-nine of the brightest stars in the area of the open cluster NGC 6250 within the Milky Way galaxy with multi-colour linear polarimetric data in five different optical bands are collected.
We then use the function $\Delta \phi$ ($\lambda$) to fit the polarized observations of each source, and apply the MHMC to simulate the linear polarization angles and the corresponding wavelengths. We find the Shapiro time delay between photons with different circular polarizations from the radio and optical candidates are in the ranges of $10^{-13}-10^{-11}$ s and $10^{-18}-10^{-15}$ s, respectively. Considering all of the values in Eq. (3), we then obtain the constraints of the $\gamma$ discrepancy with the measurements of the polarization for 60 radio sources and 29 optical stars, implying $10^{-26}-10^{-23}$ and $10^{-23}-10^{-20}$. By applying the Laniakea supercluster (instead of Milky Way) as the source of the gravitational field, the EEP constraints with the radio polarized sources can be tightened by 5-6 orders of magnitude.
We obtain the EEP constraints by using various types of sources with different polarization observations at different distances. Furthermore, the profound superiority of such tests is that nearly a few tens of sources with multi-wavelength polarization measurements are commonly available in the optical and radio bands which guarantees the significant robustness of the resulting constraint on the EEP.

\section*{Acknowledgments}
We thank Yue-Yang Zhang, Fa-Yin Wang, Ren-De Ma, Xuan Yang and Bin Liao for helpful discussion.
This work is supported by the National Natural Science Foundation
of China (Grant Nos. 11703015, U1738132, and U1831122), the Natural Science Foundation of Shandong Province (Grant No. ZR2017BA006),
the Youth Innovation Promotion Association (2017366), the Key Research Program of Frontier Sciences (grant No. ZDBS-LY-7014) of Chinese Academy of Sciences, and the Youth Innovations and Talents Project of Shandong Provincial Colleges and Universities (Grant No. 201909118).

\section*{Data availability}
The data underlying this article are available in the article.





\begin{figure*}
\includegraphics[angle=0,scale=0.30]{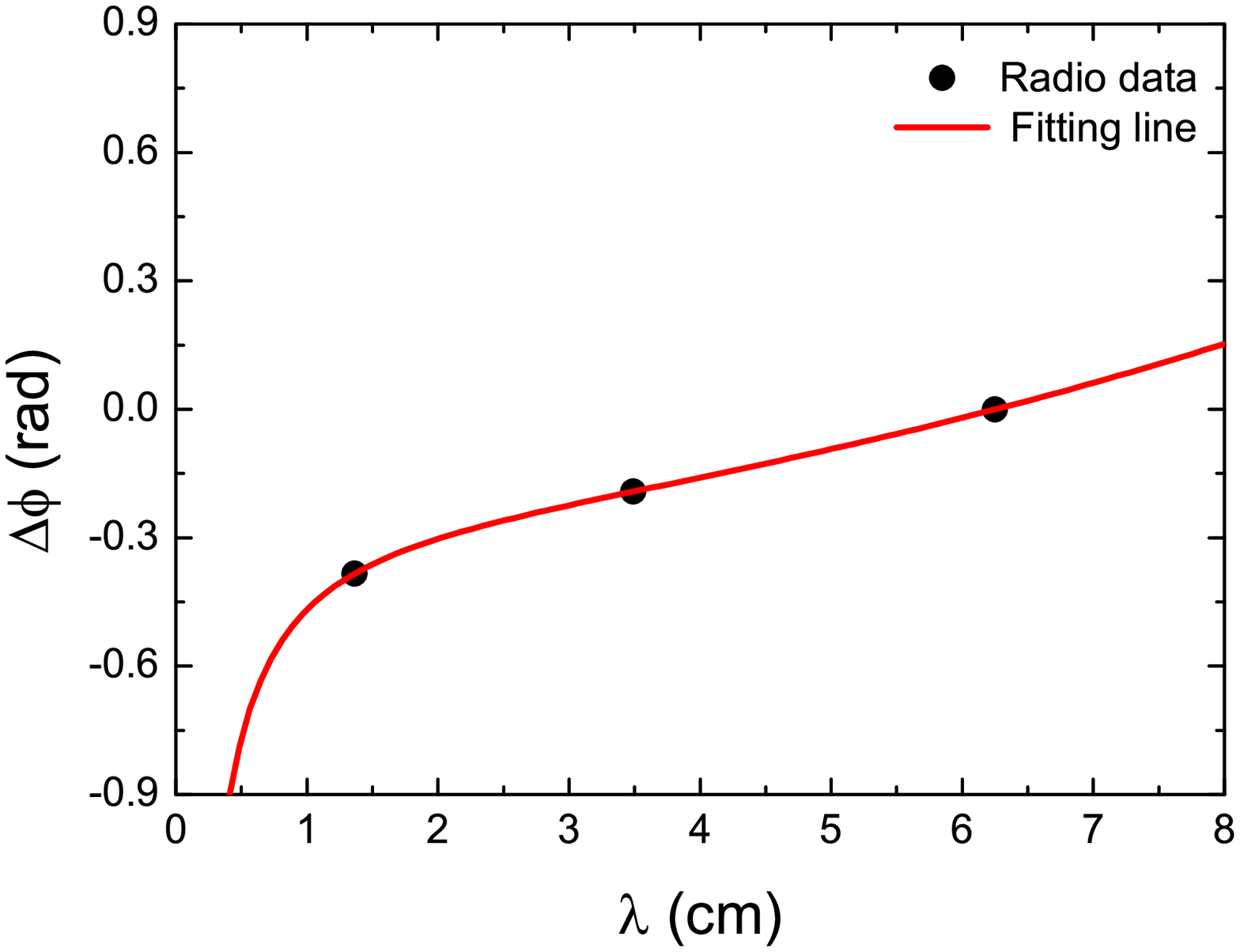}
\includegraphics[angle=0,scale=0.30]{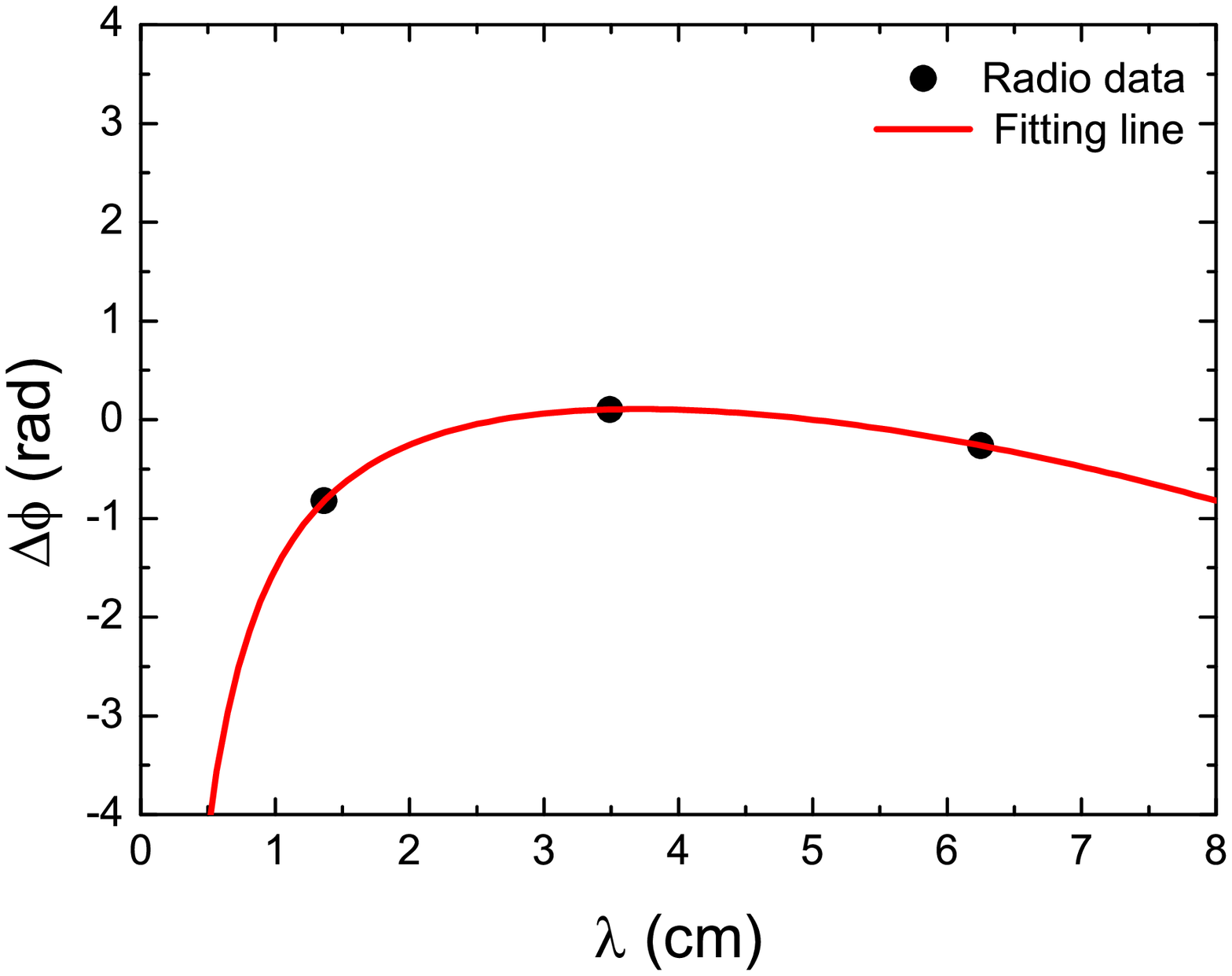}
\caption{The best-fit results of two cases with the different observed linear polarization angles and multi-wavelength observations for PKS	 0047-579 (left) and PKS 0135-247 (right).  The black dots are the linear polarization data. The red line is the fit curve using Eq. (5).}
\end{figure*}

\begin{figure*}
 \includegraphics[angle=0,scale=0.3]{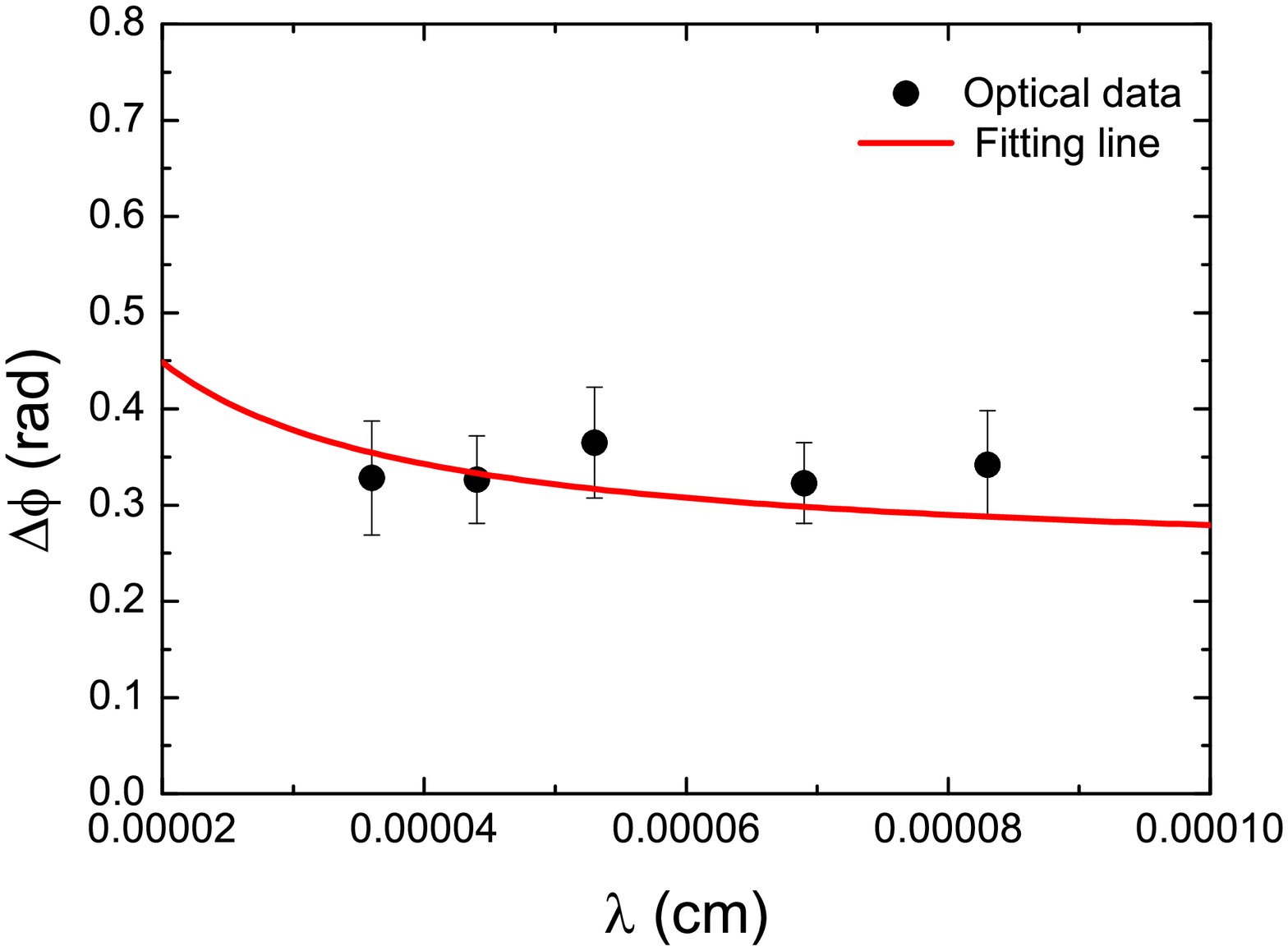}
 \includegraphics[angle=0,scale=0.3]{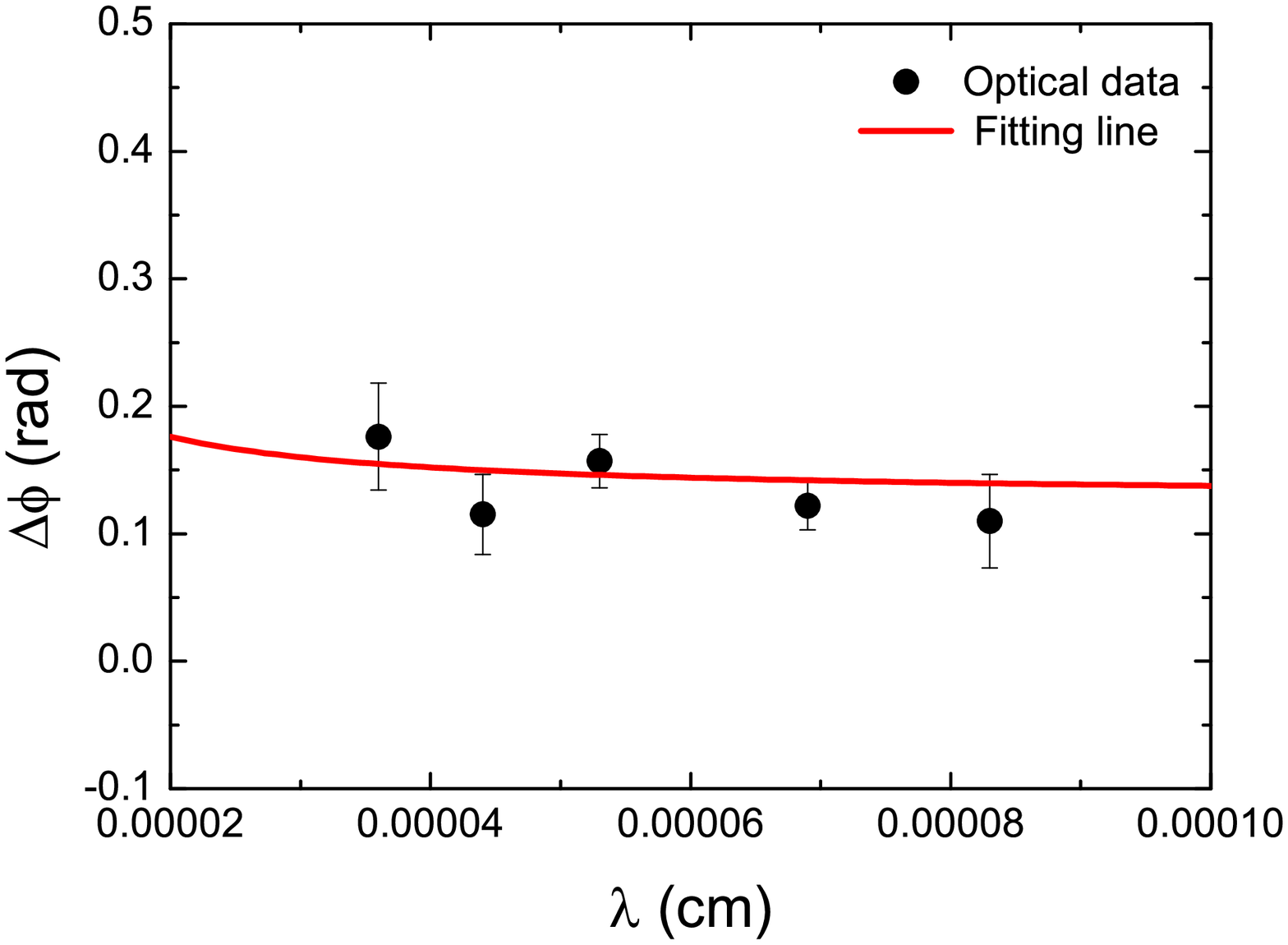}
 \caption{The best-fit results of two cases with the different observed linear polarization angles and multi-wavelength observation for optical stars in NGC 6250.   }
\end{figure*}

\begin{figure*}
 \includegraphics[angle=0,scale=0.3]{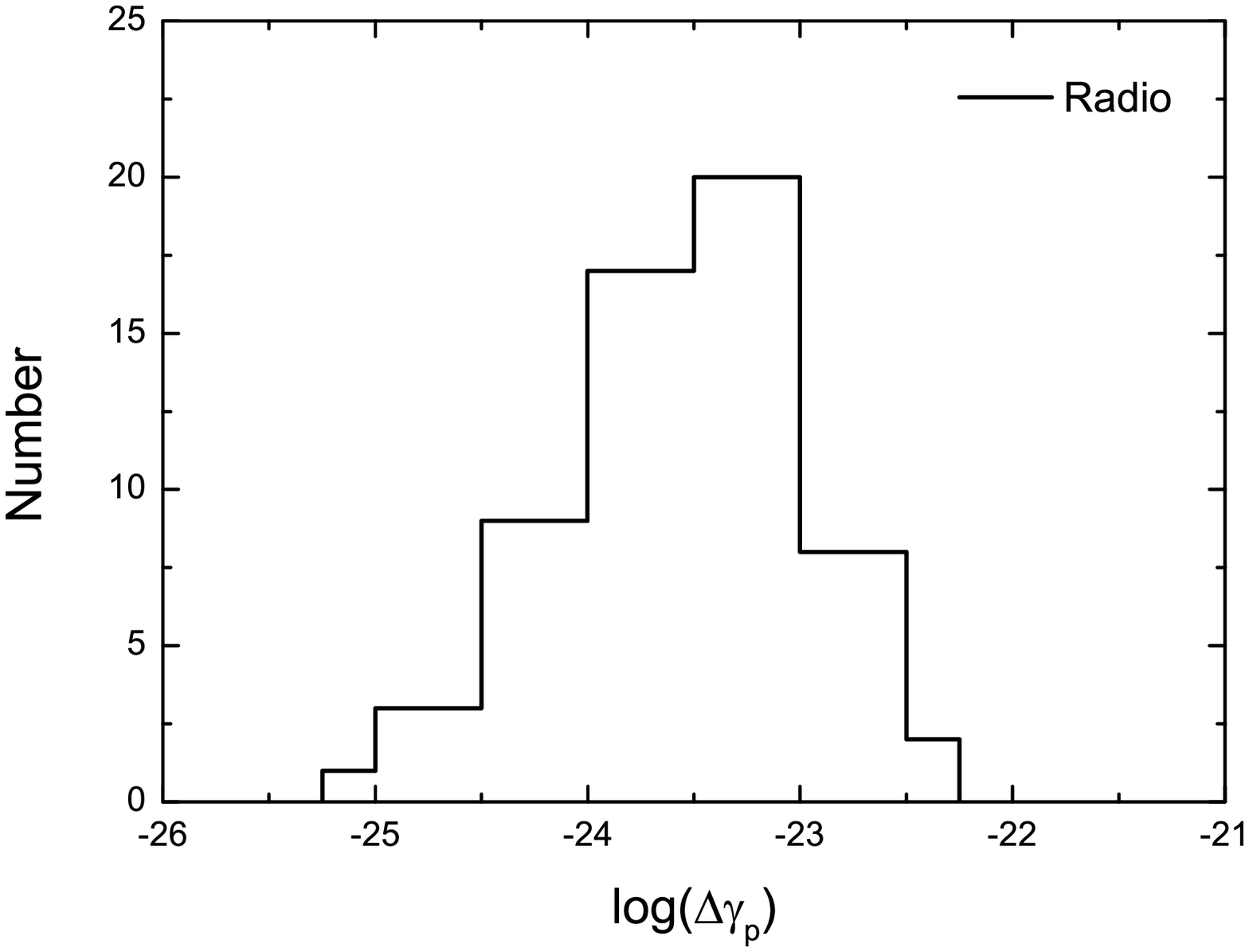}
  \includegraphics[angle=0,scale=0.3]{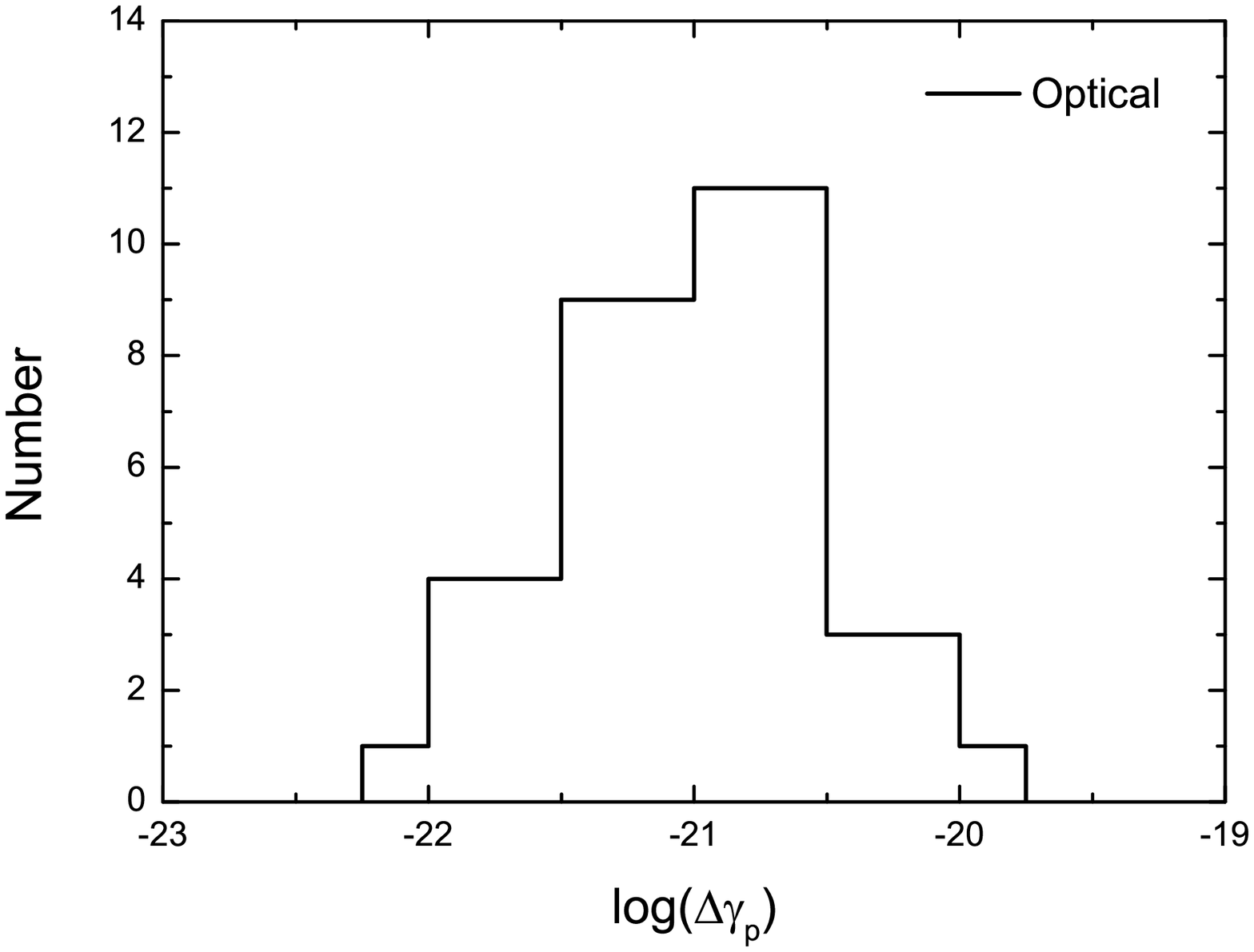}
 \caption{The distributions of the constraint results on the EEP with radio and optical candidates. }
\end{figure*}


\begin{table*}
\caption{The catalogue of 60 radio sources with multi-wavelength polarization observations. }
\label{tab:1}
\begin{tabular}{lcccccccc}
\hline
 && && 22~GHz & 43~GHz & 86~GHz &\\
\hline
Sources &RA&$\delta$ &z & $\phi_{\rm obs}(^\circ)$ & $\phi_{\rm obs}(^\circ)$ &$\phi_{\rm obs}(^\circ)$ &\\
  \hline
PKS	0047-579	&	00:49:59.48	&	-57:38:27.6	&	1.797	&	-22	&	-11	&	0	&\\
PKS	0116-219	&	01:18:57.30	&	-21:41:30.1	&	1.165	&	57	&	65	&	73	&\\
PKS	0130-17	    &	01:32:43.53	&	-16:54:48.2	&	1.02	&	34	&	39	&	48	&\\
PKS	0135-247	&	01:37:38.33	&	-24:30:53.6	&	0.837	&	-47	&	6	&	-15	&\\
PKS	0142-278	&	01:45:03.39	&	-27:33:33.9	&	1.155	&	4	&	36	&	-28	&\\
PKS	0202-17	    &	02:04:57.76	&	-17:01:20.1	&	1.74	&	-77	&	-59	&	-71	&\\
PKS	0208-512	&	02:10:46.19	&	-51:01:01.4	&	0.999	&	-7	&	-21	&	-17	&\\
PKS	0214-330	&	02:16:48.19	&	-32:47:40.6	&	1.331	&	86	&	-61	&	52	&\\
PKS	0220-349	&	02:22:56.40	&	-34:41:27.7	&	1.49	&	-26	&	-5	&	-4	&\\
PKS	0237-23	    &	02:40:08.13	&	-23:09:15.8	&	2.223	&	-33	&	-31	&	-36	&\\
PKS	0252-549	&	02:53:29.20	&	-54:41:51.4	&	0.539	&	44	&	14	&	-2	&\\
PKS	0325-222	&	03:27:59.97	&	-22:02:06.3	&	2.22	&	49	&	47	&	48	&\\
PKS	0327-241	&	03:29:54.10	&	-23:57:08.7	&	0.895	&	-62	&	22	&	37	&\\
PKS	0335-364	&	03:36:54.12	&	-36:16:06.0	&	1.541	&	16	&	8	&	43	&\\
PKS	0338-214	&	03:40:35.65	&	-21:19:30.8	&	0.223	&	8	&	0	&	-14	&\\
PKS	0346-27  	&	03:48:38.11	&	-27:49:13.4	&	0.991	&	-77	&	-67	&	0	&\\
PKS	0405-385	&	04:06:58.98	&	-38:26:27.5	&	1.285	&	14	&	3	&	-24	&\\
PKS	0405-331	&	04:07:33.92	&	-33:03:45.3	&	2.562	&	85	&	-89	&	-86	&\\
PKS	0414-189	&	04:16:36.61	&	-18:51:08.9	&	1.536	&	-85	&	-69	&	-60	&\\
PKS	0422-380	&	04:24:42.27	&	-37:56:21.0	&	0.782	&	-87	&	-71	&	87	&\\
PKS	0426-380	&	04:28:40.37	&	-37:56:19.3	&	1.11	&	-46	&	-30	&	-30	&\\
PKS	0435-300	&	04:37:36.56	&	-29:54:03.9	&	1.328	&	-66	&	-39	&	-46	&\\
PKS	0438-43	    &	04:40:17.17	&	-43:33:08.4	&	2.863	&	4	&	8	&	-65	&\\
PKS	0454-46  	&	04:55:50.79	&	-46:15:58.6	&	0.853	&	58	&	58	&	22	&\\
PKS	0454-234	&	04:57:03.23	&	-23:24:51.8	&	1.003	&	-33	&	-30	&	-44	&\\
PKS	0511-220	&	05:13:49.10	&	-21:59:17.4	&	1.296	&	-44	&	-56	&	-65	&\\
PKS	0514-459	&	05:15:45.23	&	-45:56:43.2	&	0.194	&	16	&	3	&	19	&\\
PKS	0522-611	&	05:22:34.40	&	-61:07:57.0	&	1.4	    &	47	&	65	&	82	&\\
PKS	0534-340	&	05:36:28.45	&	-34:01:10.8	&	0.684	&	-85	&	-77	&	-80	&\\
PKS	0537-441	&	05:38:50.35	&	-44:05:08.6	&	0.894	&	-44	&	-59	&	-44	&\\
PKS	0537-286	&	05:39:54.17	&	-28:39:56.3	&	3.104	&	41	&	41	&	46	&\\
PKS	0539-543	&	05:40:45.78	&	-54:18:21.7	&	1.19	&	27	&	20	&	63	&\\
PKS	0557-454	&	05:59:11.53	&	-45:29:40.4	&	0.687	&	-2	&	-70	&	-69	&\\
PKS	0607-15	    &	06:09:41.03	&	-15:42:41.6	&	0.324	&	21	&	43	&	45	&\\
PKS	0625-35	    &	06:27:06.73	&	-35:29:16.1	&	0.054	&	79	&	84	&	-84	&\\
PKS	0646-306	&	06:48:14.18	&	-30:44:19.3	&	1.153	&	-61	&	-61	&	82	&\\
PKS	0834-20	    &	08:36:39.21	&	-20:16:58.9	&	2.752	&	23	&	77	&	56	&\\
PKS	0919-260	&	09:21:29.41	&	-26:18:44.2	&	2.3	    &	18	&	7	&	1	&\\
PKS	B1102-242	&	11:04:46.06	&	-24:31:27.5	&	1.666	&	-73	&	-57	&	-77	&\\
PKS	1116-46	    &	11:18:27.08	&	-46:34:15.3	&	0.713	&	-3	&	-1	&	-6	&\\
PKS	1124-186	&	11:27:04.36	&	-18:57:19.0	&	1.05	&	30	&	46	&	36	&\\
PKS	1143-245	&	11:46:08.28	&	-24:47:34.1	&	1.94	&	-31	&	-29	&	-23	&\\
PKS	1144-379	&	11:47:01.46	&	-38:12:10.7	&	1.048	&	20	&	41	&	-73	&\\
PKS	B1206-238	&	12:09:02.64	&	-24:06:19.8	&	1.299	&	-50	&	-64	&	6	&\\
PMN	J1248-4559	&	12:48:28.53	&	-45:59:47.8	&	1.02	&	33	&	-70	&	-84	&\\
PKS	1255-316	&	12:57:59.20	&	-31:55:15.2	&	1.924	&	-76	&	-68	&	-75	&\\
PKS	B1256-177	&	12:58:38.27	&	-18:00:01.3	&	1.956	&	-15	&	-18	&	-12	&\\
PKS	1313-333	&	13:16:08.09	&	-33:38:58.9	&	1.21	&	-31	&	-2	&	-11	&\\
PKS	B1406-267	&	14:09:50.13	&	-26:57:37.3	&	2.43	&	13	&	8	&	-45	&\\
PKS	1514-24	    &	15:17:41.76	&	-24:22:20.3	&	0.048	&	54	&	44	&	36	&\\
PKS	1519-273	&	15:22:37.72	&	-27:30:11.1	&	1.294	&	-33	&	-27	&	-24	&\\
PKS	1831-711	&	18:37:28.74	&	-71:08:43.0	&	1.356	&	-69	&	-81	&	-65	&\\
PMN	J1923-2104	&	19:23:32.27	&	-21:04:33.4	&	0.874	&	-61	&	-57	&	-66	&\\
PKS	1935-692	&	19:40:25.74	&	-69:07:58.0	&	3.154	&	42	&	84	&	59	&\\
PKS	1953-325	&	19:56:59.41	&	-32:25:46.0	&	1.242	&	42	&	19	&	-10	&\\
PKS	2204-54	    &	22:07:43.82	&	-53:46:34.1	&	1.215	&	-13	&	-35	&	-47	&\\
PKS	2244-37	    &	22:47:03.81	&	-36:57:46.5	&	2.252	&	-13	&	13	&	68	&\\
PKS	2245-328	&	22:48:38.67	&	-32:35:52.5	&	2.268	&	-69	&	-82	&	88	&\\
PKS	2326-477	&	23:29:17.66	&	-47:30:19.2	&	1.299	&	-49	&	-43	&	-39	&\\
PKS	2329-384	&	23:31:59.43	&	-38:11:47.4	&	1.202	&	40	&	59	&	-88	&\\
 \hline
\end{tabular}
\end{table*}


\begin{table*}
\caption{The catalogue of 29 stars in NGC 6250 with optical polarization observations. }
\label{tab:2}
\begin{tabular}{lccccccc}
\hline
 U&B&  V & R & I &\\
\hline
$\phi_{\rm obs}(^\circ)$ & $\phi_{\rm obs}(^\circ)$ & $\phi_{\rm obs}(^\circ)$ & $\phi_{\rm obs}(^\circ)$ &$\phi_{\rm obs}(^\circ)$ &\\
  \hline
18.8	$\pm$	3.4	&	18.7	$\pm$	2.6	&	20.9	$\pm$	3.3	&	18.5	$\pm$	2.4	&	19.6	$\pm$	3.2	&\\
10.1	$\pm$	2.4	&	6.6	$\pm$	1.8	&	9	$\pm$	1.2	&	7	$\pm$	1.1	&	6.3	$\pm$	2.1	&\\
14.9	$\pm$	3.1	&	14.3	$\pm$	3.3	&	17.3	$\pm$	3.7	&	15	$\pm$	2.7	&	14.7	$\pm$	4.2	&\\
                	&	22.8	$\pm$	4.8	&	24.5	$\pm$	4.9	&	24.5	$\pm$	5.6	&	27.4	$\pm$	7.5	&\\
                 	&	19.9	$\pm$	7.2	&	24.1	$\pm$	4.8	&	22.4	$\pm$	5	&	23.4	$\pm$	4.9	&\\
13.3	$\pm$	5.6	&	22	$\pm$	3.8	&	18.4	$\pm$	5.4	&	18	$\pm$	4.8	&	4.9	$\pm$	7.5	&\\
                	&	17.3	$\pm$	2.6	&	14.6	$\pm$	1.3	&	18.9	$\pm$	1.9	&	10.9	$\pm$	2.8	&\\
15.3	$\pm$	7.6	&	15	$\pm$	3.9	&	19.7	$\pm$	5.2	&	19.3	$\pm$	3	&	12.5	$\pm$	7.1	&\\
	&	15.5	$\pm$	1.9	&	17.3	$\pm$	1	&	19	$\pm$	1.2	&	18.6	$\pm$	1.6	&\\
	&	36.1	$\pm$	3.6	&	49.7	$\pm$	3.1	&	38.8	$\pm$	3.7	&	24.9	$\pm$	6.5	&\\
	&	16	$\pm$	4.8	&	23.5	$\pm$	2.3	&	20.5	$\pm$	2	&	28.9	$\pm$	4.1	&\\
9.3	$\pm$	5.8	&	20.3	$\pm$	1.8	&	19.3	$\pm$	1.9	&	19.5	$\pm$	2.5	&	31.8	$\pm$	7.8	&\\
21.7	$\pm$	6.5	&	27	$\pm$	7	&	25.1	$\pm$	5.1	&	25.8	$\pm$	6.8	&	20.9	$\pm$	11.1	&\\
	&	44	$\pm$	4.9	&	40.1	$\pm$	5	&	44.3	$\pm$	4	&	42.4	$\pm$	4.1	&\\
12.8	$\pm$	2.9	&	14.4	$\pm$	2.2	&	16.4	$\pm$	2.5	&	12.5	$\pm$	2.1	&	17.7	$\pm$	3.7	&\\
7.2	$\pm$	4.9	&	13.7	$\pm$	3.2	&	14.8	$\pm$	3.4	&	14.7	$\pm$	1.7	&	20.3	$\pm$	3.6	&\\
25.1	$\pm$	7.4	&	27.3	$\pm$	5.9	&	41.9	$\pm$	2.9	&	7.4	$\pm$	3.9	&	13.5	$\pm$	6.6	&\\
3.8	$\pm$	3.8	&	4.8	$\pm$	3.2	&	4	$\pm$	3.2	&	4.5	$\pm$	3.2	&	6	$\pm$	4.5	&\\
12.2	$\pm$	5.4	&	4.5	$\pm$	4.1	&	7.3	$\pm$	9.3	&	12.4	$\pm$	7.7	&		&\\
44.3	$\pm$	6	&	36.8	$\pm$	5.7	&	31.6	$\pm$	3.5	&	34.5	$\pm$	2.8	&		&\\
3.5	$\pm$	7.6	&	11.9	$\pm$	1.7	&	14.1	$\pm$	4	&	13	$\pm$	3.1	&	12.5	$\pm$	4.7	&\\
	&	15	$\pm$	2.4	&	11.5	$\pm$	1.2	&	9.7	$\pm$	1.7	&	10.7	$\pm$	2.1	&\\
7.1	$\pm$	6	&	4.9	$\pm$	3.1	&	4.3	$\pm$	3.3	&	0.8	$\pm$	2	&	3.1	$\pm$	2.9	&\\
	&	35.2	$\pm$	8.7	&	42.5	$\pm$	9.7	&	46.7	$\pm$	6.7	&	44.3	$\pm$	9.1	&\\
31.7	$\pm$	4.6	&	31.8	$\pm$	4.6	&	28.7	$\pm$	8.9	&		&	14.9	$\pm$	9.5	&\\
178.3	$\pm$	5.6	&	175	$\pm$	3	&	1.2	$\pm$	5.6	&	4.2	$\pm$	5.6	&	3.2	$\pm$	7.6	&\\
26.6	$\pm$	2.5	&	29.9	$\pm$	1.1	&	31.7	$\pm$	1.2	&	30.2	$\pm$	1	&	28.4	$\pm$	1.6	&\\
14.4	$\pm$	5.7	&	27.3	$\pm$	2.9	&	27.8	$\pm$	2.8	&	30.8	$\pm$	2.7	&	33.6	$\pm$	3.8	&\\
	&	20.2	$\pm$	1.6	&	23	$\pm$	0.6	&	21	$\pm$	0.4	&	17.7	$\pm$	0.5	&\\
 \hline
\end{tabular}
\end{table*}


\begin{table*}
\caption{Best-fit results of multi-frequency polarization observations for 60 radio sources.}
\label{tab:2}
\begin{tabular}{lcccccc}
\hline
 Sources & $\phi_{\rm 0} \,(rad)$ &A (cm) & B ($cm^{-2}$) & $\Delta t_{gra}$~($\times 10^{-12}s$)&$\Delta \gamma_{p}$~($\times 10^{-24}$) &\\
  \hline
PKS	0047-579	&	-0.177	&	-0.297	&	0.00574	&	-3.16	&	1.03	&\\		
PKS	0116-219	&	1.15	&	-0.216	&	0.00417	&	-2.29	&	0.807	&\\		
PKS	0130-17	&	0.633	&	-0.0680	&	0.00552	&	-0.721	&	0.256	&\\		
PKS	0135-247	&	1.19	&	-2.68	&	-0.0262	&	-28.4	&	10.9	&\\		
PKS	0142-278	&	1.99	&	-2.48	&	-0.0532	&	-26.3	&	9.51	&\\		
PKS	0202-17	&	-0.594	&	-0.991	&	-0.0125	&	-10.5	&	3.35	&\\		
PKS	0208-512	&	-0.632	&	0.681	&	0.00580	&	7.23	&	2.74	&\\		
PKS	0214-330	&	-4.83	&	8.34	&	0.113	&	88.5	&	31.6	&\\		
PKS	0220-349	&	0.216	&	-0.904	&	-0.00361	&	-9.59	&	3.37	&\\		
PKS	0237-23	&	-0.442	&	-0.172	&	-0.00405	&	-1.82	&	0.568	&\\		
PKS	0252-549	&	0.0108	&	1.05	&	-0.00546	&	11.1	&	4.82	&\\		
PKS	0325-222	&	0.776	&	0.105	&	0.00114	&	1.11	&	0.345	&\\		
PKS	0327-241	&	1.45	&	-3.43	&	-0.00642	&	-36.4	&	13.8	&\\		
PKS	0335-364	&	-0.460	&	0.939	&	0.0271	&	9.96	&	3.49	&\\		
PKS	0338-214	&	0.0708	&	0.115	&	-0.00854	&	1.23	&	0.660	&\\		
PKS	0346-27	&	-1.93	&	0.688	&	0.0467	&	7.30	&	2.74	&\\		
PKS	0405-385	&	0.256	&	0.0285	&	-0.0174	&	0.302	&	0.111	&\\		
PKS	0405-331	&	-4.22	&	7.68	&	0.0381	&	81.5	&	25.6	&\\		
PKS	0414-189	&	-1.09	&	-0.550	&	0.00325	&	-5.84	&	1.93	&\\		
PKS	0422-380	&	-3.16	&	1.95	&	0.112	&	20.7	&	8.54	&\\		
PKS	0426-380	&	-0.282	&	-0.701	&	-0.00330	&	-7.44	&	2.82	&\\		
PKS	0435-300	&	-0.178	&	-1.30	&	-0.0107	&	-13.8	&	4.90	&\\		
PKS	0438-43	&	1.20	&	-1.40	&	-0.0540	&	-14.9	&	4.65	&\\		
PKS	0454-46	&	1.50	&	-0.605	&	-0.0262	&	-6.42	&	2.57	&\\		
PKS	0454-234	&	-0.287	&	-0.367	&	-0.0108	&	-3.89	&	1.43	&\\		
PKS	0511-220	&	-1.04	&	0.375	&	-0.00408	&	3.98	&	1.38	&\\		
PKS	0514-459	&	-0.363	&	0.839	&	0.0143	&	8.90	&	5.43	&\\		
PKS	0522-611	&	1.17	&	-0.503	&	0.00866	&	-5.34	&	1.81	&\\		
PKS	0534-340	&	-1.18	&	-0.401	&	-0.00384	&	-4.26	&	1.79	&\\		
PKS	0537-441	&	-1.46	&	0.910	&	0.0140	&	9.65	&	3.84	&\\		
PKS	0537-286	&	0.647	&	0.0840	&	0.00364	&	0.892	&	0.267	&\\		
PKS	0539-543	&	-0.345	&	1.03	&	0.0328	&	10.9	&	3.92	&\\		
PKS	0557-454	&	-2.26	&	3.00	&	0.0148	&	31.8	&	13.4	&\\		
PKS	0607-15	&	1.05	&	-0.931	&	-0.00309	&	-9.88	&	4.64	&\\		
PKS	0625-35	&	3.84	&	-3.04	&	-0.123	&	-32.3	&	34.0	&\\		
PKS	0646-306	&	-3.02	&	2.40	&	0.104	&	25.5	&	9.38	&\\		
PKS	0834-20	&	2.45	&	-2.72	&	-0.0264	&	-28.9	&	8.65	&\\		
PKS	0919-260	&	0.0384	&	0.381	&	-0.00210	&	4.05	&	1.28	&\\		
PKS	B1102-242	&	-0.480	&	-1.04	&	-0.0179	&	-11.0	&	3.68	&\\		
PKS	1116-46	&	0.0811	&	-0.172	&	-0.00405	&	-1.82	&	0.755	&\\		
PKS	1124-186	&	1.18	&	-0.869	&	-0.0106	&	-9.23	&	3.34	&\\		
PKS	1143-245	&	-0.558	&	0.0132	&	0.00396	&	0.140	&	0.0456	&\\		
PKS	1144-379	&	2.59	&	-2.84	&	-0.0873	&	-30.1	&	11.7	&\\		
PKS	B1206-238	&	-2.29	&	1.79	&	0.0539	&	19.0	&	6.70	&\\		
PMN	J1248-4559	&	-2.58	&	4.28	&	0.0111	&	45.4	&	17.3	&\\		
PKS	1255-316	&	-0.970	&	-0.468	&	-0.00675	&	-4.97	&	1.66	&\\		
PKS	B1256-177	&	-0.441	&	0.232	&	0.00499	&	2.47	&	0.786	&\\		
PKS	1313-333	&	0.526	&	-1.42	&	-0.0125	&	-15.1	&	5.59	&\\		
PKS	B1406-267	&	0.789	&	-0.672	&	-0.0376	&	-7.12	&	2.24	&\\		
PKS	1514-24	&	0.727	&	0.304	&	-0.00376	&	3.23	&	3.36	&\\		
PKS	1519-273	&	-0.422	&	-0.213	&	9.45E-04	&	-2.26	&	0.810	&\\		
PKS	1831-711	&	-1.81	&	0.795	&	0.0141	&	8.44	&	2.74	&\\		
PMN	J1923-2104	&	-0.811	&	-0.327	&	-0.00738	&	-3.47	&	1.34	&\\		
PKS	1935-692	&	2.44	&	-2.26	&	-0.0269	&	-24.0	&	6.73	&\\		
PKS	1953-325	&	0.381	&	0.521	&	-0.0164	&	5.53	&	2.04	&\\		
PKS	2204-54	&	-0.779	&	0.763	&	-0.00419	&	8.09	&	2.86	&\\		
PKS	2244-37	&	-0.133	&	-0.216	&	0.0347	&	-2.29	&	0.749	&\\		
PKS	2245-328	&	-3.95	&	3.43	&	0.126	&	36.4	&	11.9	&\\		
PKS	2326-477	&	-0.715	&	-0.196	&	0.00167	&	-2.08	&	0.738	&\\		
PKS	2329-384	&	3.33	&	-3.30	&	-0.111	&	-35.0	&	13.0	&\\		
 \hline
\end{tabular}
\end{table*}


\begin{table*}
\caption{Best-fit results of 29 bright stars with optical polarization observations.}
\label{tab:4}
\begin{tabular}{lcccccc}
\hline
 $\phi_{\rm 0} \,(rad)$ &A (cm) & B ($cm^{-2}$) & $\Delta t_{gra}$~($\times 10^{-17}s$)&$\Delta \gamma_{p}$~($\times 10^{-22}$) &\\
  \hline
0.237	$\pm$	0.0464	&	4.23	$\pm$	2.38	&	0.131	$\pm$	0.0270	&	4.49	$\pm$	2.52	&	9.61	$\pm$	 5.41	&\\
0.128	$\pm$	0.0369	&	0.962	$\pm$	1.87	&	0.137	$\pm$	0.0473	&	1.02	$\pm$	1.98	&	2.19	$\pm$	 4.25	&\\
0.245	$\pm$	0.0826	&	0.855	$\pm$	3.69	&	-0.115	$\pm$	0.0373	&	0.907	$\pm$	3.92	&	1.94	$\pm$	 8.40	&\\
0.285	$\pm$	0.102	&	6.59	$\pm$	5.81	&	0.150	$\pm$	0.0348	&	6.99	$\pm$	6.16	&	15.0	$\pm$	 13.2	&\\
0.320	$\pm$	0.0222	&	4.00	$\pm$	3.20	&	0.00344	$\pm$	0.0714	&	4.25	$\pm$	3.40	&	9.10	$\pm$	 7.29	&\\
0.472	$\pm$	0.0365	&	-6.20	$\pm$	2.58	&	-0.133	$\pm$	0.0497	&	-6.57	$\pm$	2.74	&	14.1	$\pm$	 5.87	&\\
0.313	$\pm$	0.0303	&	-1.81	$\pm$	1.88	&	-0.00545	$\pm$	0.0211	&	-1.92	$\pm$	1.99	&	4.11	 $\pm$	 4.27	 &\\
0.411	$\pm$	0.0502	&	-5.57	$\pm$	3.14	&	0.0622	$\pm$	0.0419	&	-5.91	$\pm$	3.34	&	12.7	$\pm$	 7.15	&\\
0.245	$\pm$	0.0625	&	3.30	$\pm$	3.49	&	-0.154	$\pm$	0.0275	&	3.50	$\pm$	3.70	&	7.51	$\pm$	 7.93	&\\
0.201	$\pm$	0.0474	&	27.8	$\pm$	3.08	&	-0.125	$\pm$	0.0253	&	29.5	$\pm$	3.27	&	63.2	$\pm$	 7.00	&\\
0.252	$\pm$	0.0377	&	6.85	$\pm$	2.63	&	0.0899	$\pm$	0.0437	&	7.27	$\pm$	2.79	&	15.6	$\pm$	 5.97	&\\
0.208	$\pm$	0.0230	&	5.95	$\pm$	1.49	&	-0.00968	$\pm$	0.0551	&	6.31	$\pm$	1.58	&	13.5	 $\pm$	 3.39	 &\\
0.412	$\pm$	0.0802	&	0.556	$\pm$	4.39	&	0.113	$\pm$	0.0424	&	0.590	$\pm$	4.66	&	1.26	$\pm$	 9.99	&\\
0.664	$\pm$	0.0557	&	4.59	$\pm$	3.87	&	0.0385	$\pm$	0.0694	&	4.87	$\pm$	4.11	&	10.4	$\pm$	 8.80	&\\
0.247	$\pm$	0.0469	&	0.252	$\pm$	2.33	&	0.115	$\pm$	0.0586	&	0.268	$\pm$	2.47	&	0.574	$\pm$	 5.29	&\\
0.402	$\pm$	0.0563	&	-8.06	$\pm$	2.97	&	-0.111	$\pm$	0.0652	&	-8.55	$\pm$	3.15	&	18.3	$\pm$	 6.75	&\\
0.597	$\pm$	0.0852	&	2.01	$\pm$	4.59	&	0.156	$\pm$	0.0315	&	2.13	$\pm$	4.87	&	4.57	$\pm$	 10.4	&\\
0.0538	$\pm$	0.0265	&	0.863	$\pm$	1.89	&	-0.0767	$\pm$	0.0659	&	0.915	$\pm$	2.00	&	1.96	$\pm$	 4.29	&\\
0.0631	$\pm$	0.0324	&	2.91	$\pm$	2.62	&	-0.0456	$\pm$	0.0655	&	3.08	$\pm$	2.78	&	6.61	$\pm$	 5.96	&\\
0.0754	$\pm$	0.0363	&	25.1	$\pm$	2.71	&	0.0688	$\pm$	0.0322	&	26.7	$\pm$	2.87	&	57.2	$\pm$	 6.16	&\\
0.265	$\pm$	0.0292	&	-2.67	$\pm$	1.74	&	-0.0739	$\pm$	0.0417	&	-2.84	$\pm$	1.84	&	6.08	$\pm$	 3.95	&\\
0.132	$\pm$	0.0289	&	3.90	$\pm$	1.75	&	0.0710	$\pm$	0.0343	&	4.14	$\pm$	1.86	&	8.86	$\pm$	 3.99	&\\
0.218	$\pm$	0.0359	&	-5.88	$\pm$	2.31	&	-0.0781	$\pm$	0.0460	&	-6.23	$\pm$	2.46	&	13.4	$\pm$	 5.26	&\\
0.913	$\pm$	0.0496	&	-10.2	$\pm$	5.19	&	-0.0122	$\pm$	0.112	&	-10.8	$\pm$	5.50	&	23.2	$\pm$	 11.8	&\\
0.0692	$\pm$	0.0429	&	19.3	$\pm$	2.74	&	0.0362	$\pm$	0.0381	&	20.5	$\pm$	2.91	&	43.8	$\pm$	 6.24	&\\
0.0209	$\pm$	0.0180	&	111	$\pm$	1.99	&	0.173	$\pm$	0.0281	&	118	$\pm$	2.11	&	253	$\pm$	4.52	&\\
0.698	$\pm$	0.0451	&	-7.80	$\pm$	2.14	&	0.0651	$\pm$	0.0140	&	-8.28	$\pm$	2.27	&	17.7	$\pm$	 4.86	&\\
0.558	$\pm$	0.0586	&	-5.05	$\pm$	3.12	&	0.0126	$\pm$	0.0401	&	-5.36	$\pm$	3.31	&	11.5	$\pm$	 7.09	&\\
0.318	$\pm$	0.0232	&	3.64	$\pm$	1.45	&	-0.170	$\pm$	0.0177	&	3.86	$\pm$	1.53	&	8.27	$\pm$	 3.29	&\\
 \hline
\end{tabular}
\end{table*}

\end{document}